\documentclass[twocolumn,aps,prl]{revtex4}
\usepackage{graphicx}
\usepackage{dcolumn}
\usepackage{amsmath}
\usepackage{latexsym}
\usepackage{verbatim}

\begin {document}

\title {Nonstationary but quasisteady states in Self-organized Criticality}
\author
{S. S. Manna}
\affiliation
{
\begin {tabular}{c}
Satyendra Nath Bose National Centre for Basic Sciences,
Block-JD, Sector-III, Salt Lake, Kolkata-700106, India
\end{tabular}
}
\begin{abstract}

      The notion of Self-organized criticality (SOC) had been conceived to interpret the spontaneous emergence
   of long range correlations in nature. Since then many different models had been introduced to study SOC. All 
   of them have few common features: externally driven dynamical systems self-organize themselves to non-equilibrium 
   stationary states exhibiting fluctuations of all length scales as the signatures of criticality. In contrast, 
   we have studied here in the framework of the sandpile model a system that has mass inflow but no outflow. 
   There is no boundary, and particles cannot escape from the system by any means. Therefore, there is no current 
   balance, and consequently it is not expected that the system would arrive at a stationary state. In spite of 
   that, it is observed that the bulk of the system self-organizes to a quasi-steady state where the grain density 
   is maintained at a nearly constant value. Power law distributed fluctuations of all length and time scales have 
   been observed which are the signatures of criticality. Our detailed computer simulation study gives the set of 
   critical exponents whose values are very close to their counter parts in the original sandpile model. This study 
   indicates that (i) a physical boundary and (ii) the stationary state though sufficient but may not be the necessary
   criteria for achieving SOC.

\end{abstract}

\maketitle

      An externally driven, non-linear system with open boundary are the key points in the prescription of a        
   self-organized critical system \cite {BTW}. The driving instrument adds intermittently mass (or energy) to 
   the system in the form of tiny particles. The dynamics is non-linear since the rule does not allow the local 
   accumulation of particles indefinitely \cite {Jensen,Frontiers,Hoffmann,Garber}. This is incorporated using a cut-off in the particle number, beyond 
   which the particles get distributed. This way the system responds to the external drive to minimize the effect 
   of the drive that creates inhomogeneity in particle density. The mass distribution takes place in a `domino' 
   process, creates a series of activity in the form of an avalanche. Eventually, all these activities subside 
   due to spreading of particles through the self-organizing diffusion process and also by flushing out of particles
   from the system across the boundary. The system is continued to be driven ever after, repeatedly \cite {Dhar,Dhar1,Ram,Dickman}.

      Thus, in their original prescription \cite {BTW} Bak et. al. designed such a nonequilibrium system 
   with a steady inflow of mass through the driving process and outflow through the boundary. As a result, a 
   stationary state sets in when these two currents balance each other. In this state the avalanches in the 
   system are observed to be of all length and time scales, which are considered to be the signatures of the 
   long-range spatio-temporal correlations and appearance of the critical state in the system
   \cite {Levine,Manna}.

      It was claimed that the steady flow of particle current through the system and the settling of the system
   in a stationary state are the necessary conditions to achieve the SOC state. On a careful look however, one
   realizes, since the ratio of the numbers of boundary to bulk sites becomes very small in the limit of 
   asymptotically large systems, there may be little effect of the boundary in this problem. It had been observed
   that indeed an increasing number of avalanches remain confined to the bulk as the system size become larger which are 
   not touched by the boundary. This is because the probability distribution of linear extent (diameter) of the 
   avalanches are also observed to decay as a power law \cite {Satya}.

      This observation leads us to argue that presence of a physical boundary and establishing a stationary state 
   under the external drive may not be the absolutely necessary criteria to achieve self-organized criticality. 
   In particular, no current balance to attain the stationary state is really required. In the following we would
   describe that even on an infinite system without a boundary the system can infact self-organize to a nearly steady 
   state. We devise a model system using the frameworks of the well known sandpile models of SOC \cite {Dhar2,Dhar3,
   Manna1,Manna2} where such a non-stationary but quasi-steady state in the bulk is produced.

%---------------------------------------------------------------------------------
\begin{figure}[t]
\includegraphics[width=6.0cm]{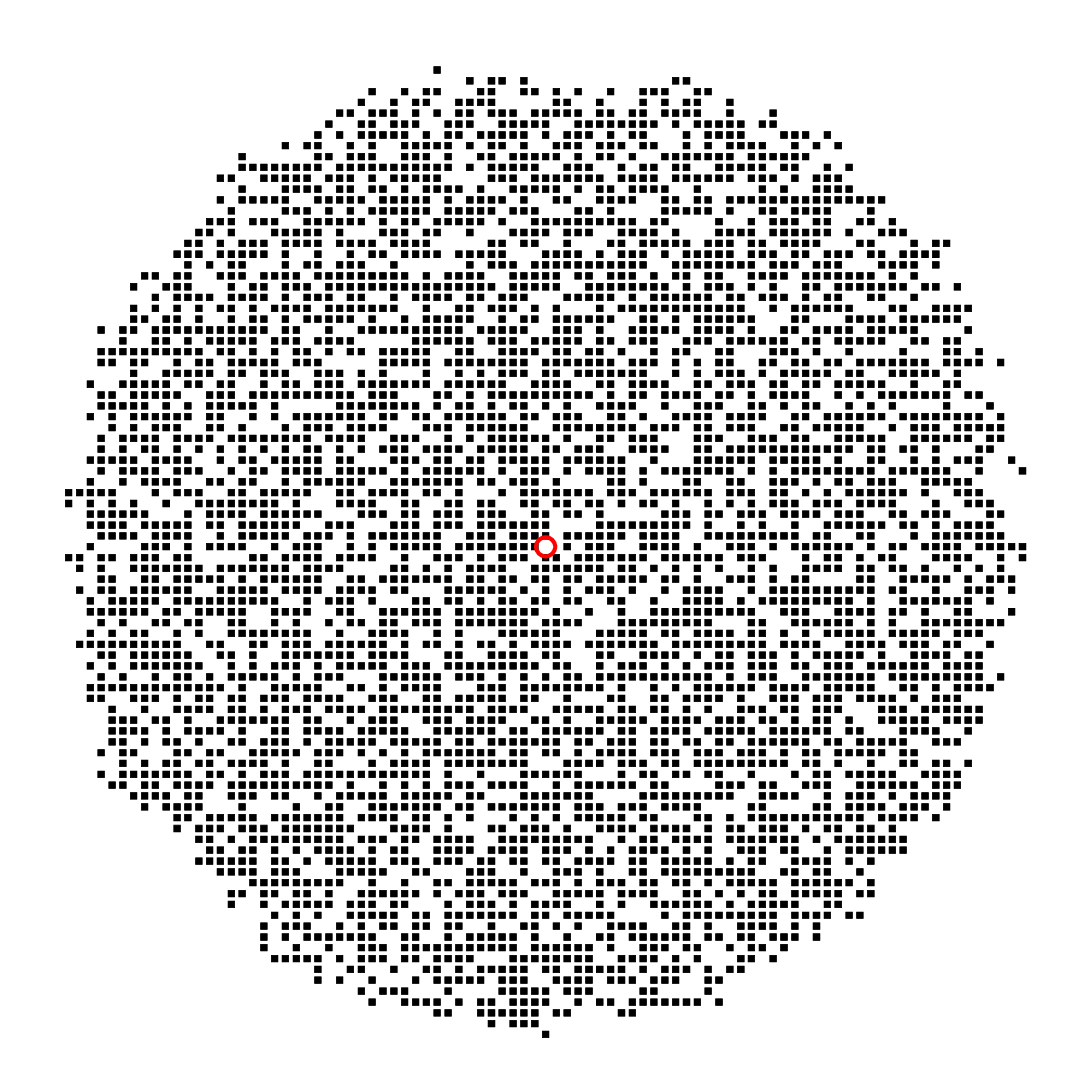} \\
\caption{
      At the origin (marked by the red circle) of the infinite square lattice $N$ = 4096 particles have been dropped 
   one by one. The distribution of particles (black square symbols) in the final passive state has been shown. Apart 
   from a thin outer interface, the bulk of the system has nearly a constant density of particles.
   }
\label {FIG01}
\end{figure}
%---------------------------------------------------------------------------------

      We construct a growing sandpile on an infinite square lattice which we consider as the $x-y$ plane. Sand particles are 
   dropped one by one only at the origin of the coordinate system. When a particle is dropped, some activity is generated in the 
   system through the hard core collision process following the dynamical rule of the non-abelian stochastic sandpile model 
   \cite {Manna}. A collision is said to take place if more than one particle share a lattice site at the same time when each
   particle selects one neighbour site randomly with uniform probability and moves there. As time evolves, the number of collisions initially grows, 
   reaches a maximum and then goes down and eventually this activity dies after some time. Such a state is referred as the 
   `passive state' when no particle moves. The next particle is then dropped again at the origin. Therefore, the addition of a single 
   sand particle takes the system from one passive state to another passive state through a sequence of activities. This entire set 
   of activities together is called an `avalanche'. Different avalanches create different impacts to the system and their strengths 
   are measured by the sizes of the avalanches. Most commonly, the size $s$ of an avalanche is measured by the total number of 
   collisions that take place in the entire avalanche. 

%---------------------------------------------------------------------------------
\begin{figure}[t]
\includegraphics[width=5.5cm]{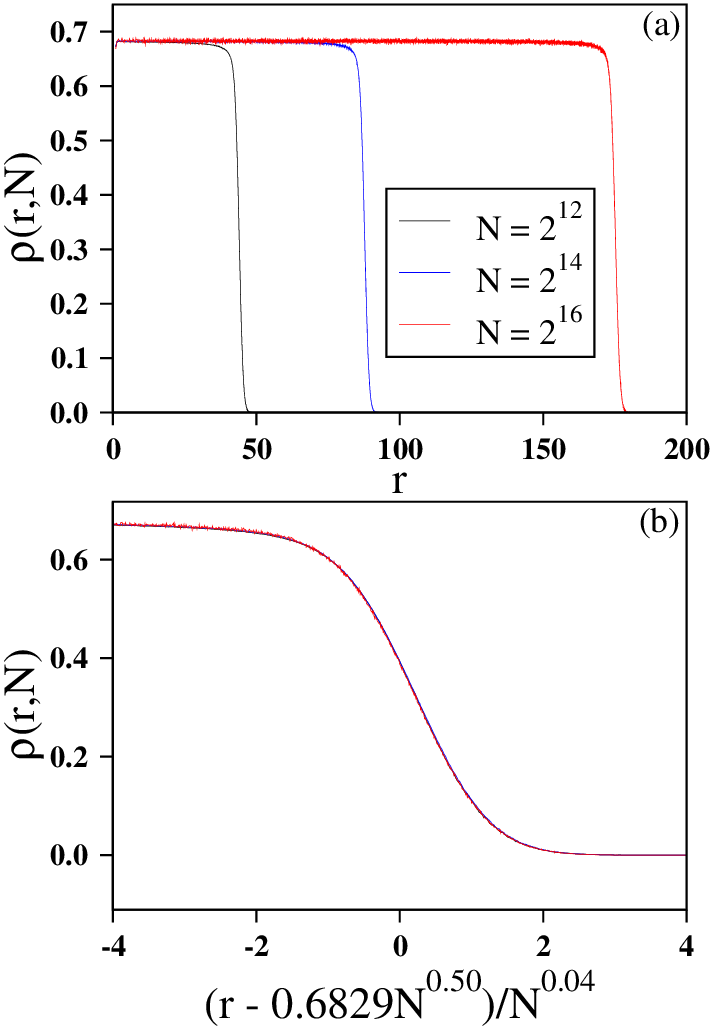} \\
\caption{
   (a) The particle density $\rho(r,N)$ in the passive state at a distance $r$ from the origin have been plotted
   after dropping $N$ particles one by one at the origin.
   (b) The same data of (a) have been plotted again after scaling the $x$-axis by $(r - 0.6829N^{0.50})/N^{0.04}$.
   The three plots collapse nicely on top of one another. 
   }
\label {FIG02}
\end{figure}
%---------------------------------------------------------------------------------

      In the passive state, a lattice site is either occupied by one particle or it is vacant. Typically occupied sites are randomly 
   distributed on the lattice (Fig. \ref {FIG01}). Sometimes the origin may also be occupied by one particle. Therefore, when a particle 
   is dropped at the origin, it is likely that a collision takes place which then triggers an avalanche. If some of the neighboring sites are 
   also occupied there would be further collisions at these sites and a cascade of collisions results.

      We first characterize the passive state by the variation of particle density after dropping $N$ particles one at a time
   at the origin. The density $\rho(r,N)$ is the average number of particles at a site located
   at a distance $r$ from the origin. When we plot $\rho(r,N)$ against $r$ in Fig. \ref {FIG02}(a) for the three different
   values of the total number $N$ = $2^{12}$, $2^{14}$, and $2^{16}$ of particles dropped, we observe a flat bulk region for all $N$. 
   In this region, the particle densities are nearly the same though there is a very small but systematic $N$ dependence.
   We find the average bulk density $\rho(N) = \rho(\infty) - AN^{-x_1}$ where $\rho(\infty)$ = 0.6835 and $x_1$ = 0.484 are found.
   For this reason we say the bulk of the system has reached the quasi-steady state.

%---------------------------------------------------------------------------------
\begin{figure}[t]
\includegraphics[width=8.5cm]{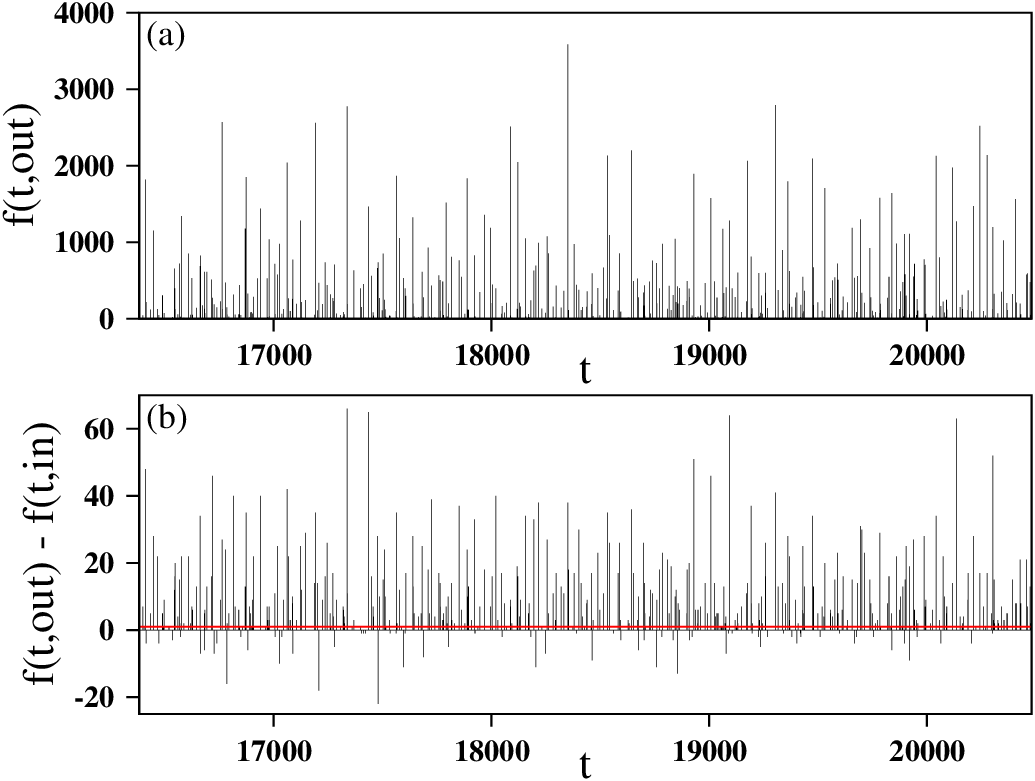} \\
\caption{
   (a) The out flux $f(t,out)$ through a circle of radius $R = 32$ centered at the origin per particle addition has been 
   plotted against time $t$ between $N$ and $N + \Delta N$ with $N = 2^{14}$ and $\Delta N = 2^{12}$. For each avalanche 
   there is an influx $f(t,in)$ in general which found to be of the same order as $f(t,out)$ and therefore, it has not 
   been plotted.
   (b) The net outward flux $f(t,out) - f(t,in)$ has been plotted. It is observed that most of the time it is positive 
   i.e., outward, whereas less frequently it is inward as well. The red line shows the average net flux on this interval 
   which is very close to unity due to unit rate of particle addition.
   }
\label {FIG03}
\end{figure}
%---------------------------------------------------------------------------------

      As the distance $r$ from the origin increases, the bulk region is followed by an interface where the particle density gradually goes
   down and finally vanishes. This interface shifts to larger $r$ values as $N$ increases. The steepness of the fall of density profile 
   increases on increasing $N$. We define a half radius $r_{1/2}(N)$ where the density is half of its average bulk value $\rho(N)$.
   On plotting (not shown) $r_{1/2}(N)$ against $N$ on a double logarithmic scale we find $r_{1/2}(N) = 0.6829N^{0.50}$ on the average.
   We use it in Fig. \ref {FIG02}(b) for a scale transformation $r - r_{1/2}(N)$. It makes all three curves pass through nearly
   the same point, but their slopes at this point are different. To make them collapse on one another we have to scale the $x$-axis 
   by $N^{-0.04}$. Therefore, we finally plot again $\rho(r,N)$ against $(r - 0.6829N^{0.50})/N^{0.04}$ to obtain a nice data collapse. 

      The next question we ask is how the particle density in the bulk is maintained as the system evolves? Other than a constant inflow 
   of particles at the origin, and since no particle goes out of the system by evaporation or by other means, it is a fully conservative 
   system. These particles only get themselves distributed to the larger space through the diffusive collision process but they maintain 
   the bulk density and consequently the interface of the particle system moves outwards.

%---------------------------------------------------------------------------------
\begin{figure}[t]
\includegraphics[width=5.5cm]{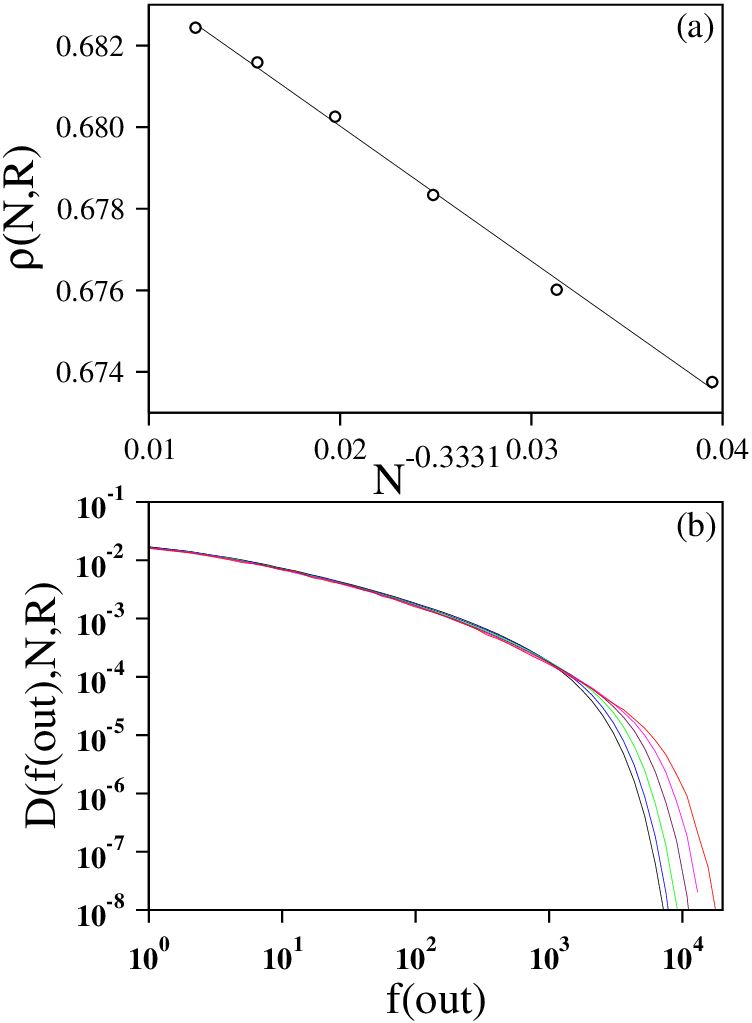} \\
\caption{
   (a) The average particle density $\rho(N,R)$ within the circle ${\cal C}$ of radius $R$ measured within the time interval
   between $N$ and $N+\Delta N$ has been plotted against $N^{-0.3331}$ and is extrapolated to $\rho(\infty,R)$ = 0.6866.
   (b) Probability distribution of $D(f(out),N,R)$ of the outward flux $f(out)$ across the same circle for every particle 
   addition at the origin within the time interval between $N$ and $N+\Delta N$ has been plotted against $f(out)$. When 
   $N$ is increased from left to right, the distribution is enlarged and larger values of the outward fluxes become more 
   probable. In both plots $N = 2^{14}, 2^{15}, ..., 2^{19}$ and $\Delta N = 2^{15}$.
   }
\label {FIG04}
\end{figure}
%---------------------------------------------------------------------------------

      To see this point in more detail we consider a circle ${\cal C}$ of radius $R$ centered at the origin, situated well inside the bulk region 
   created by dropping $N$ particles. Now we continue to drop another $\Delta N$ particles at the origin, one at a time again, and observe the 
   flux of the particle current through the circle. There would be collisions at sites both inside and outside that are adjacent to the 
   circle. Therefore, for each particle addition at the origin, some particles would cross the circle from inside to outside, where as 
   some other particles would come to inside from outside. In actual simulation, we mark all sites inside the circle and keep the outer 
   sites as unmarked. Corresponding to each avalanche, we count how many particles jumped from marked to unmarked sites which
   constitute the flux of outflow current $f(t,out)$. Similarly, the number of particles that jumped from unmarked to marked sites 
   constitute flux of the inflow current $f(t,in)$.
   
      In Fig. \ref {FIG03}(a) we have shown the variation of $f(t,out)$ for $\Delta N = 4096$ time units after dropping $N = 2^{14}$
   particles. In almost all avalanches $f(t,in)$ is smaller than $f(t,out)$ but of the same order, occasionally however
   $f(t,in)$ is larger. Therefore, we do not plot the variation of $f(t,in)$ which almost look the same, but plot of the net flux
   $f(t,out) - f(t,in)$ against time in Fig. \ref {FIG03}(b) which is mostly positive, but sometimes negative too. The average
   net flux $\langle f(t,out) - f(t,in) \rangle$ over the entire interval is very close to its exact value unity and has been 
   marked using the red line.

%---------------------------------------------------------------------------------
\begin{figure}[t]
\includegraphics[width=5.5cm]{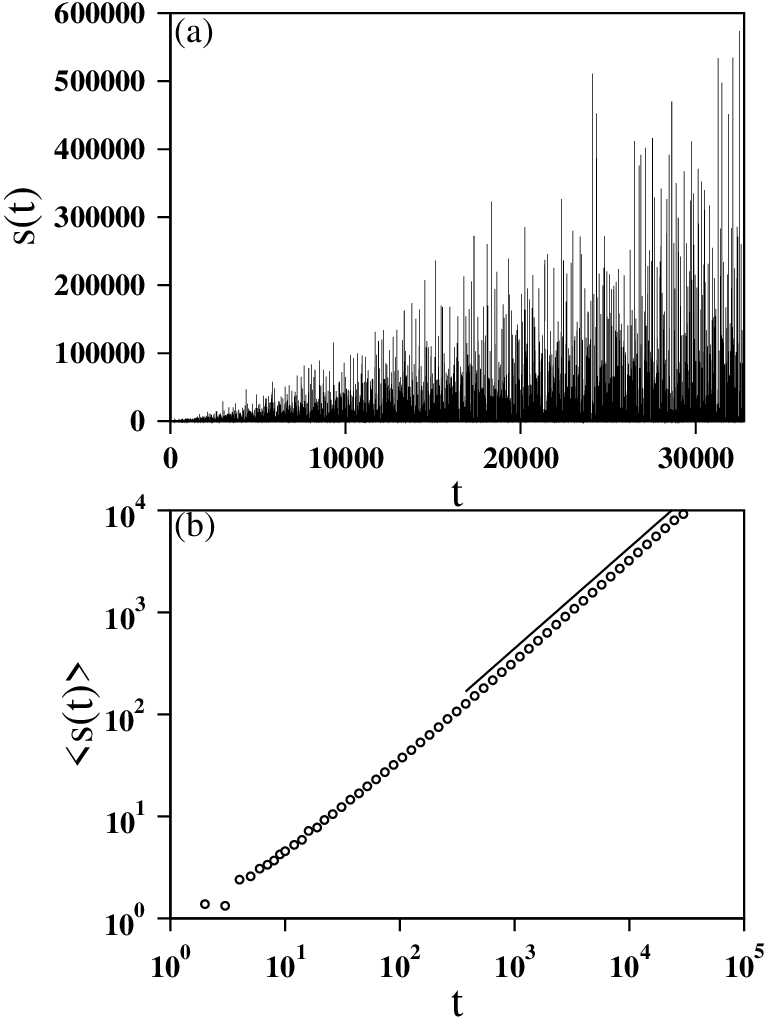} \\
\caption{
   (a) Plot of size $s(t)$ of the avalanche created by dropping the $t$-th particle at the origin for a single run
   of $N = 2^{15}$ particles.
   (b) The average value of the avalanche size $\langle s(t) \rangle$ has been plotted against time $t$ on a 
   $\log - \log$ scale, the slope of the straight line fitting the curve is 0.988.
   }
\label {FIG05}
\end{figure}
%---------------------------------------------------------------------------------

      Similarly, for any finite volume within the bulk region like the circle ${\cal C}$ the flux of outflow and inflow currents must balance
   on the average. No particle leaves the system on infinite lattice, the system only self-organizes itself and particles spread out through 
   the collision process. Within the circle ${\cal C}$ the system tries to achieve the steady state of constant density which it never 
   succeeds in finite time, it only approaches its asymptotic value as $N$ increases. In Fig. \ref {FIG04}(a) we have plotted $\rho(N,R)$
   which is the average particle density within ${\cal C}$ during the time interval $N$ to $N+\Delta N$. On extrapolation it gives the density
   0.6866 in the asymptotic limit of $N \to \infty$. This shows that even the core of the bulk region has not become completely steady
   but it has attained a quasi-steady state and slowly approaches its asymptotic state.

      A similar conclusion can also be drawn by looking at the probability distribution $D(f(out),N,R)$ of the outward fluxes $f(out)$
   from the same circle ${\cal C}$ calculated within the time interval $\Delta N$ after first skipping an initial time $N$ (Fig. \ref {FIG04}(b)). It 
   has been observed that on increasing $N$ the distribution shifts to the larger out flux regime and there is no trace of the
   distribution reaching a steady time independent form. This study shows that the bulk of the system does not reach a true steady state 
   in finite time but slowly approaches its quasi-steady form.

%---------------------------------------------------------------------------------
\begin{figure}[t]
\includegraphics[width=5.5cm]{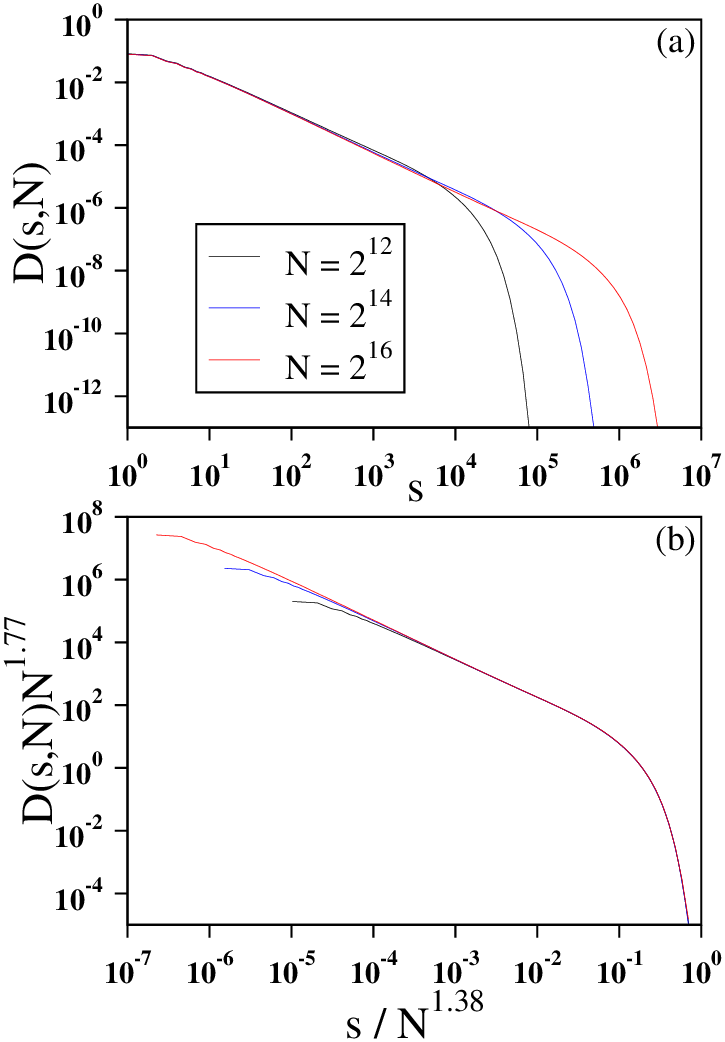} \\
\caption{
   (a) The probability distributions $D(s,N)$ of the sizes of avalanches have been plotted 
   against the avalanche size $s$ for $N$ particles dropped one by one. The data have been collected 
   over one million runs in each case.
   (b) The same data of (a) have been plotted again after scaling $D(s,N)N^{1.77}$ against $s / N^{1.38}$
   yielding the value of the avalanche size exponent $\tau = 1.77 / 1.38 \approx 1.28$.
   }
\label {FIG06}
\end{figure}
%---------------------------------------------------------------------------------

      Now we would like to explore, if this self-organized state is critical or not. For that we have to check if there are
   fluctuations of all length scales present in the system. Accordingly, we have defined the avalanche sizes $s$ and life
   times $T$ in the following way. When a particle is dropped at the origin, it creates a sequence of activities in the system. 
   The lattice sites where particle collisions take place are updated synchronously. Let the intra-avalanche time be denoted by 
   $T$. The list of collision sites at time $T$ are updated to create the same list at time $T+1$. The avalanche is finished 
   when the length of the list shrinks to zero. The total number of time steps $T$ is the life-time of the avalanche and the 
   total number of collisions $s$ is the avalanche size. Initially, their magnitudes are very small but they gradually grow 
   and soon become quite large. In Fig. \ref {FIG04}(a) we have shown the variation of the size $s(t)$ of the avalanche created 
   by dropping the $t$-th particle at the origin. The time series is for a single run when a total of $N=2^{15}$ particles have been 
   dropped. Next we average the avalanche size over many different runs and plot the average avalanche size $\langle s(t) \rangle$ 
   against $t$ on a double logarithmic scale in Fig. \ref {FIG04}(b). The slope of the curve for large time is found to be 0.988 
   which indicates that the avalanche size possibly increases linearly with time.

%---------------------------------------------------------------------------------
\begin{figure}[t]
\includegraphics[width=5.5cm]{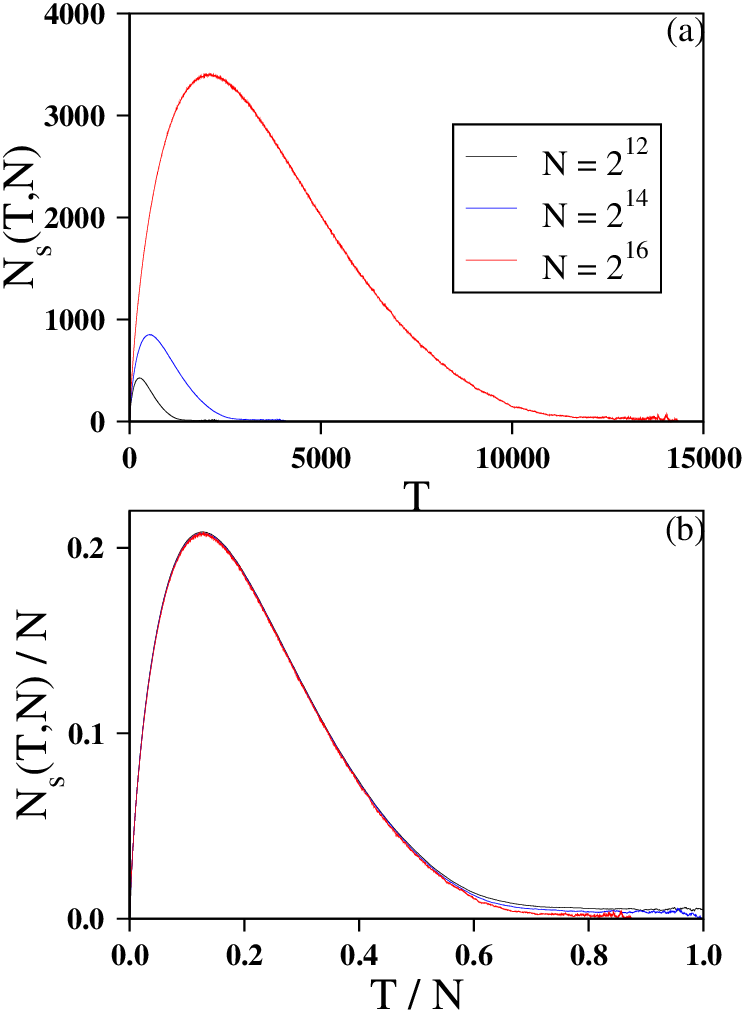} \\
\caption{
   Relaxation of $N$ particles dropped simultaneously at the origin at time $T = 0$.
   (a) The number $N_s(T,N)$ of active sites at the intra-avalanche time $T$ first sharply increases to a maximum 
   and then gradually decreases to zero. 
   (b) The same data of (a) have been plotted again after scaling the axes by $N$ to obtain a nice data collapse. The scaled 
   curve fits to the generalized Gamma distribution.
   }
\label {FIG07}
\end{figure}
%---------------------------------------------------------------------------------

      We have calculated the probability distributions $D(s,N)$ and $D(T,N)$ of the sizes and life-times of the 
   avalanches respectively when a total of $N$ particles are dropped. In Fig. \ref {FIG06}(a) we have plotted $D(s,N)$ 
   against $s$ for three different $N$ values, namely $2^{12}$, $2^{14}$ and $2^{16}$, the avalanche size data 
   have been collected over many independent runs. The plots exhibited the characteristics of power law distributions 
   measured in finite systems. On the double logarithmic scale they have the linear region in the middle leading 
   to a bending and sharp fall at some high value cut-off size $s_c(N)$. The linear regions have slopes $\approx 1.200, 
   1.226$ and 1.231 respectively for the three $N$ values. The cut-off size increases with $N$ by approximately the 
   same amount on the log-log scale when $N$ is increased by the same factor. This implies $s_c(N) \sim N^{\alpha}$, 
   where $\alpha$ is the scaling exponent to be determined. Further, we have done a finite-size scaling analysis in 
   Fig. \ref {FIG06}(b). We observe that the distributions $D(s,N)$ scales nicely using suitable powers of $N$, like:
\begin {equation}
D(s,N)N^{\beta} \propto {\cal G}(s/N^{\alpha})
\end {equation}
   where, ${\cal G}(x)$ is the scaling function such that ${\cal G}(x) \rightarrow x^{-\tau}$ for
   $x << 1$ and ${\cal G}(x) \rightarrow constant$ for $x >> 1$. The limiting distribution
   $D(s) = \lim_{N\to\infty}D(s,N) \propto s^{-\tau}$ must be independent of $N$ which leads
   to $\tau = \beta/\alpha$. To try this finite size scaling analysis we have scaled the $x$-axis
   by $s/N^{1.38}$ and the $y$-axis as $D(s,N)N^{1.77}$. We have tuned the scaling exponents and selected these values for
   the best fit. Therefore, this scaling analysis gives $\tau = 1.77/1.38 \approx 1.28$.
   A similar analysis for the life-times distribution yields
\begin {equation}
D(T,N)N^{\beta_T} \propto {\cal G_T}(T/N^{\alpha_T})
\end {equation}
   where $\alpha_T = 0.77$ and $\beta_T = 1.15$ and $\tau_T = 1.494$.
   The average values of avalanche size and life-times are found to grow like: $\langle s(N) \rangle \sim N$ and 
   $\langle T(N) \rangle \sim N^{0.41}$.

%---------------------------------------------------------------------------------
\begin{figure}[t]
\includegraphics[width=5.5cm]{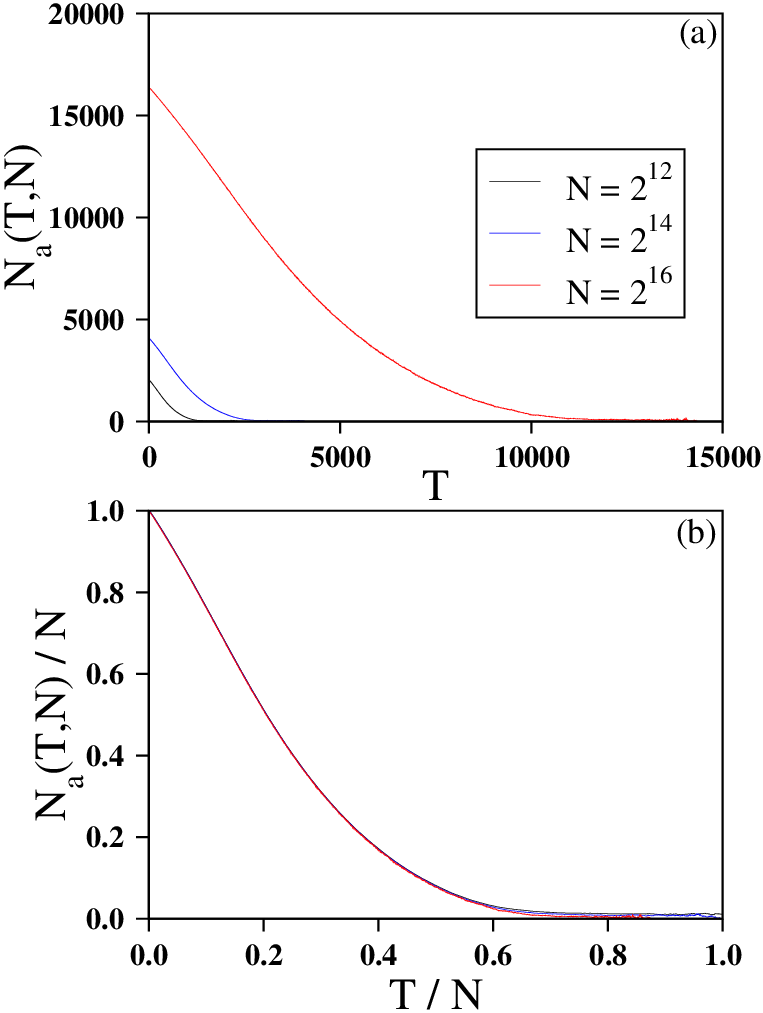} \\
\caption{
   Relaxation of $N$ particles dropped simultaneously at the origin at time $T = 0$.
   (a) The number $N_a(T,N)$ of active particles at the intra-avalanche time $T$ initially sharply decreases
   and then slowly vanishes leading to the passive state.
   (b) The same data of (a) have been plotted again after scaling both the axes by $N$ to obtain a nice 
   data collapse. The scaled curves fit nicely to the shifted Gaussian (Eqn. 4).
   }
\label {FIG08}
\end{figure}
%---------------------------------------------------------------------------------

      To check if these results are consistent with the model of ordinary finite size sandpile 
   we have studied the same system having fixed open boundary. On an $L \times L$ system, the particles
   have been dropped one by one only at the centre of the lattice. In this model, the collisions which take place on the
   boundary may throw particles outside the system if these directions are randomly selected. Consequently,
   the stationary state corresponds to the balance of inflow and outflow currents of sand particles. The avalanche
   sizes have been measured for different system sizes $L$, namely, 65, 129, and 257. Again a data collapse analysis
   has turned out to be very successful when $D(s,L)L^{3.46}$ have been plotted against $s/L^{2.72}$ (figure not shown).
   This implies the exponent $\tau$ is $3.46 / 2.72 \approx 1.272$ which matches very well with the value 1.28 of the
   same exponent in the infinite system.     

      In the original sandpile model \cite {Manna} the steady state is robust with respect to the choice of the initial state
   to start with which is the signature of the self-organizing dynamical process. Consequently,
   the particle density in the steady state is independent of the density of particles in the initial state.
   Here also we see the same phenomenon. In another study we add all $N$ particles together at the origin. When 
   such a system evolves to the passive state we find the density profile indisinguishable 
   from the density profile of the first version when particles were dropped one by one at the origin. Therefore, 
   the nearly same bulk density for all three $N$ values exhibits the signature of self-organization by the dynamical 
   process of this model. These results indicate that even without using a fixed boundary for the mass outflow and current balance the system 
   can achieve the self-organized state.

%---------------------------------------------------------------------------------
\begin{figure}[t]
\begin {center}
\includegraphics[width=8.5cm]{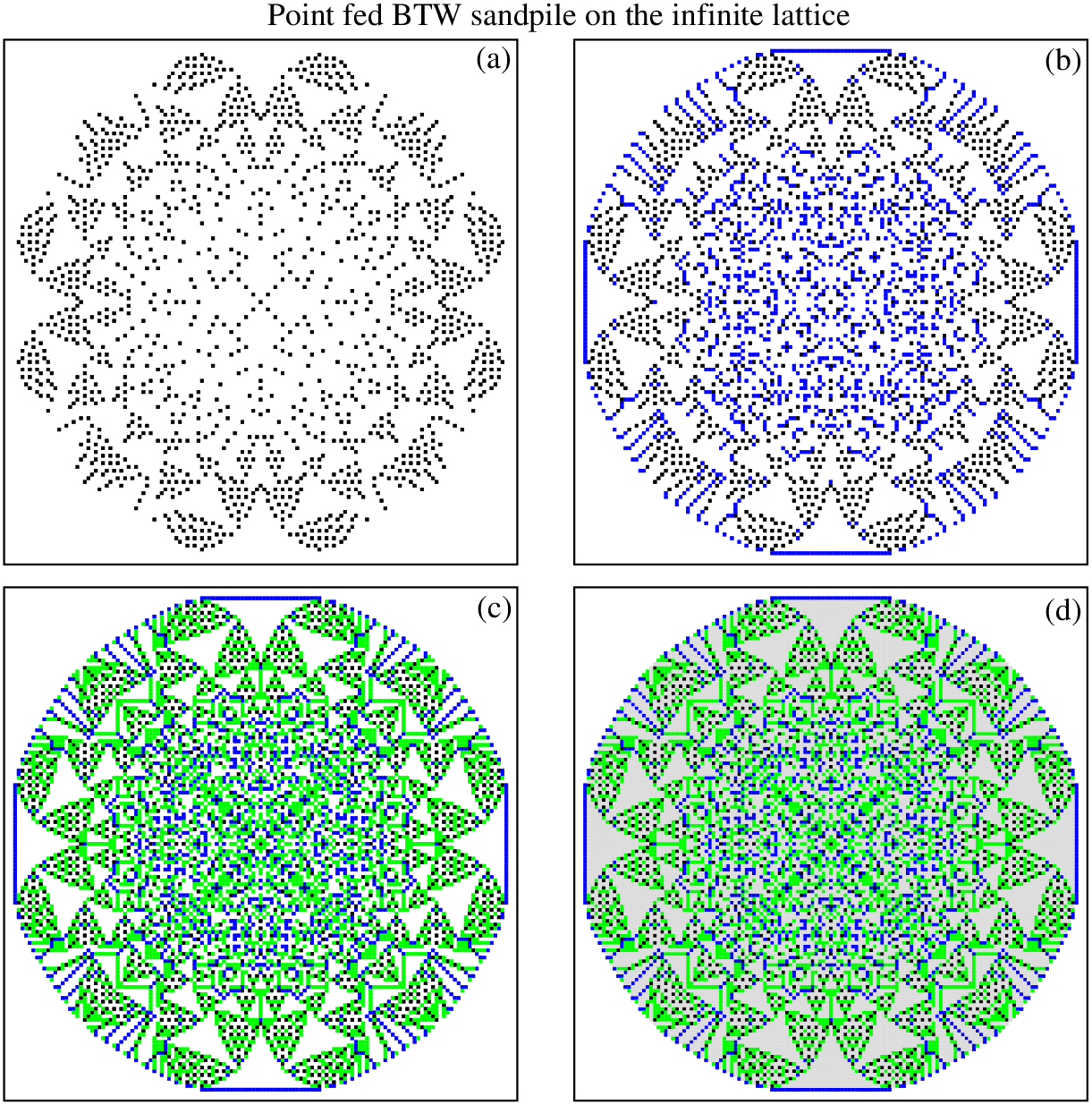} \\
\end {center}
\caption{
   A total of $N=2^{15}$ particles have been dropped one by one only at the origin of the infinite square lattice. The final 
   stable state height configuration has been drawn representing sand column heights 0, 1, 2, 
   and 3 by color dots: black (1477), blue (1532), green (4032), and gray (7724) respectively. 
   (a) Sites of height 0 only, (b) sites with heights 0 and 1, (c) sites with heights 0, 1, and 2, and finally
   (d) sites with heights 0, 1, 2, and 3.
   }
\label {FIG09}
\end{figure}
%---------------------------------------------------------------------------------

      Next we have studied how this system evolves starting from such an initial condition. Specifically, after 
   adding $N$ particles at time $T = 0$ at the origin of the infinite square lattice and we study how the system 
   relaxes as the time $T$ increases. At time $T$ = 1, each particle jumps to one of the neighbouring sites selecting it randomly. 
   In the next time step they again jump to their nearest neighbours. In general, collision dynamics is followed, 
   i.e., at any intermediate time if there are more than one particle at a site at a time, then all particles randomly
   jump to the neighbouring sites. As before, this dynamics stops only when there is no active site present in the system, i.e., 
   the system reaches a passive state.

      Two quantities are measured. At any arbitrary intermediate time $T$ we counted the number 
   $N_s(T,N)$ of active sites, i.e. sites which have more than one particle at that time. It is 
   observed that $N_s(T,N)$ first grows, reaches a maximum and then decays to a passive state 
   when there is no activity at all. The number of such active sites have been averaged
   over a large number of independent runs to obtain $\langle N_s(T,N) \rangle$. In Fig. 
   \ref {FIG07}(a) we have plotted $\langle N_s(T,N) \rangle$ against $T$ for $N = 2^{12}$, $2^{14}$, 
   and $2^{16}$. As $N$ becomes larger, the height of the peak as well as the duration of the
   avalanche increases. In the next Fig. \ref {FIG07}(b) we plot again the same data but after scaling 
   the axes. We have plotted $\langle N_s(T,N) \rangle /N$ against $T/N$ and get a nice data collapse.
   To find its functional form we find that the scaled curve fits best to a generalized Gamma distribution function:
\begin{equation}
   y =a_0(x/a_1)^{\zeta}\exp(-(x/a_1)^{\eta}).
\end{equation}
   where, $y = \langle N_s(T,N) \rangle /N$, $x = T / N$ and the best fitted parameters are: $a_0 = 0.456$, 
   $a_1 = 0.234$, $\eta = 0.63$ and $\zeta = 1.484$.

%---------------------------------------------------------------------------------
\begin{figure}[t]
\includegraphics[width=5.5cm]{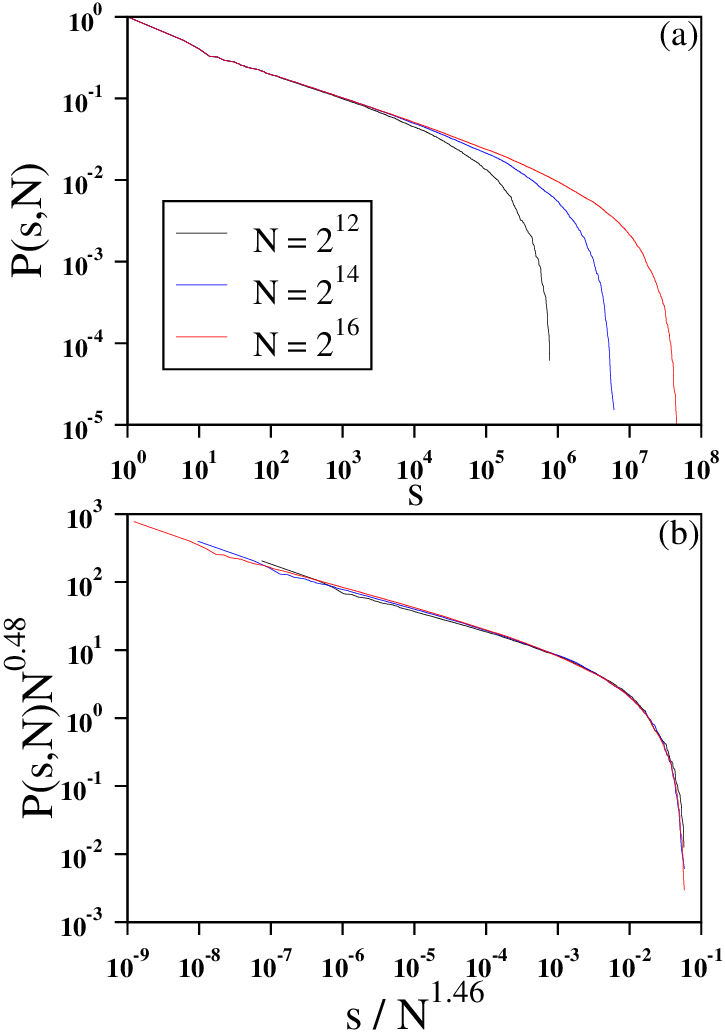} \\
\caption{
   The cumulative probability distribution $P(s,N)$ of the avalanche sizes of the BTW model
   where sand particles have been dropped only at the origin. (a) The distribution $P(s,N)$ have
   been plotted against the avalanche size $s$ for three different values of $N$.
   (b) A nice data collapse of the same data is observed when we have plotted $P(s,N)N^{0.48}$
   against $s / N^{1.46}$. This analysis implies that the avalanche size exponent $\tau = 1+ 0.48/1.46 \approx 1.328$.
   }
\label {FIG10}
\end{figure}
%---------------------------------------------------------------------------------

      Secondly, we have measured how the number of active particles $N_a(T,N)$ decays with time $T$ and finally vanishes.
   As time progresses the particles spread to a larger region, where they hardly get other particles to collide and therefore
   become more and more inactive. In Fig. \ref {FIG08}(a) we show the plot of the average number of active particles    
   $\langle N_a(T,N) \rangle$ against $T$ for $N = 2^{12}$, $2^{14}$, and $2^{16}$. Initially each curve decays fast, 
   but then it slows down and finally vanishes when the passive state is reached. In the next plot Fig. \ref {FIG08}(b) 
   we scale the axes and plot $\langle N_a(T,N) / N \rangle$ against $T / N$ which again gave a nice collapse of the data. 
   The best fitted form of this collapsed plot is the shifted Gaussian:
\begin{equation}
   y =a_0\exp(-a_1(x+a_2)^2).
\end{equation}
   where, $y = \langle N_a(T,N) / N \rangle$, $x = T / N$ and the best fitted parameters for $N = 2^{16}$ are $a_0 = 1.21$
   which is decreasing towards unity on increasing $N$, $a_1 = 5.94$ which is increasing and $a_2 = 0.18$ which is also
   gradually decreasing to zero.

      Next we perform a similar study for the deterministic Bak, Tang, and Wiesenfeld (BTW) sandpile model 
   \cite {BTW,Wiesenfeld} on the 
   infinite square lattice and as before we drop sand particles one by one only at the origin. As per
   the rule of the BTW sand pile, the sand column of height $h(i,j)$ becomes unstable only
   when it exceeds a predefined height $z - 1$. An unstable sand column topples and redistributes
   sand particles as:
\begin {center}
\begin {tabular}{c}
\verb !h(i,j)! $\rightarrow$ \verb !h(i,j) - z! \\
\verb !h(i!$\pm1$, \verb!j!$\pm1$) $\rightarrow$ \verb!h(i!$\pm1$, \verb!j!$\pm1$) + 1
\end {tabular}
\end {center}
   and $z = 4$ is chosen for the square lattice.

      We first notice that since the dynamics is entirely deterministic, the underlying
   symmetries of the square lattice determine the particle distribution patterns. In Fig. \ref {FIG09}
   the particle distribution patterns have been displayed after dropping $N = 2^{15}$ particles one by one at
   the origin. The final stable configuration has four fold symmetry. For clarity we have shown
   four figures with sites of heights (a) 0 only, (b) 0 and 1, (c) 0, 1, and 2, (d) 0, 1, 2, and 3. 

      Since after dropping every four particles the origin becomes active, therefore there are a total of $N/4$
   avalanches when $N$ particles are dropped. As more and more particles are dropped the avalanche sizes become
   gradually larger. On calculating the probability distribution of the avalanche sizes $D(s,N)$
   we find the data to be very much fluctuating. Therefore, we consider the cumulative probability distribution
   $P(s,N)$ i.e., the probability that a randomly selected avalanche has size $s$ or larger. This integrated
   distribution is much smoother as displayed in Fig. \ref {FIG10}(a) for $N = 2^{12}, 2^{14}$, and $2^{16}$. 
   In addition, we execute a finite size scaling here as well. In Fig. \ref {FIG10}(b) we have plotted $P(s,N)N^{0.48}$
   against $s/N^{1.46}$ and obtain a nice collapse of the data. This implies that the avalanche size exponent
   $\tau = 1+0.48/1.46 \approx 1.33$.

      To check if this avalanche size exponent matches with the same system but with a boundary, we have studied
   the same BTW model on an $L \times L$ square lattice having the centre at the origin. As in the ordinary BTW model
   particles dropped outside the boundary. Only difference here being the system is fed by dropping particles at the origin 
   only. In the stationary state, avalanche sizes are measured for a long time and the cumulative probability distribution
   $P(s,L)$ has been calculated for different $L$ = 65, 129, and 257. A finite size scaling plot of $P(s,L)L$ against
   $s/L^{2.9}$ exhibited a good data collapse (not shown here), yielding $\tau = 1+1/2.9 \approx 1.34$. Therefore the 
   avalanche size distribution exponents for the single site fed BTW model on a square lattice with or without boundaries 
   very well match (1.33 against 1.34) to each other.

      Finally, for the same system we have estimated the probability $P_L$ that an arbitrary avalanche reaches the boundary
   in the steady state starting from the center of the $L \times L$ lattice. That means $P_L$ is the fraction of avalanches
   that dropped at least one particle outside the system. Our numerical estimation gives $P_L \sim L^{-0.82}$. We recognize
   this exponent to be the same as the cumulative probability distribution $P(\xi) \sim \xi^{1-\tau_{\xi}}$  of the linear
   extent $\xi$ of the avalanches and therefore, $\tau_{\xi} \approx 1.82$ \cite {Satya}.

      To summarize, we have found a way to generate the self-organized critical state without a physical boundary. In the
   original models of SOC the physical entity, mass or energy for example, can drop out of the system through such a boundary. Here
   we have studied a growing sandpile where particles have been injected one by one at the origin of the infinite square lattice. 
   Addition of each particle created an avalanche of activities in the system which eventually dies down and the system 
   returns to a new passive state. This passive state is not only self-organized but also critical since it exhibits 
   long range correlations of all length scales. Since there is no boundary, the data are found to be much well behaved.

      In contrast to the original prescription of the Bak, Tang, and Wiesenfeld \cite {BTW} we observe that a steady 
   flow of particle current through the system where the average fluxes of global inflow and the global outflow balance 
   each other may not be an absolutely necessary criterion. Instead, only the external drive that injects an inflow 
   current so that the particles only get scattered within the system as per the dynamical rules of the model is sufficient 
   to ensure the self-organized criticality in the system. It is also true that the balance of the outward flux and inward 
   flux of particles through any arbitrarily defined fixed volume within the bulk of the system is always maintained. 
   Because of the particle number conservation and the self-organizing dynamical process a quasi-steady particle density in 
   the bulk is maintained.

      Similar study with abelian stochastic sandpile where only two particles are transferred in a collision is under progress.

      I thank very much one of the referees who suggested the study of the outflux of particles through a box in the bulk of the
   system. Also acknowledge that a substantial part of the numerical work has been done in S. N. Bose National Centre for Basic Sciences, 
   Kolkata through a Visiting (Honorary) Fellow position till 31-st January 2023.

\end {document}